\documentclass[conference]{IEEEtran}
\IEEEoverridecommandlockouts
\usepackage[linesnumbered,ruled,noend]{algorithm2e}
\newboolean{techreport}
\setboolean{techreport}{true}
\usepackage{graphicx}
\usepackage{balance}
\usepackage{bbm}
\usepackage{url}
\usepackage{cite}
\usepackage{amsmath}
\usepackage{amsfonts}
\usepackage{subcaption}
\usepackage{amssymb}
\usepackage{times}
\usepackage[table,xcdraw,prologue]{xcolor}
\usepackage{cleveref}
\usepackage{multirow}
\usepackage{xcolor}
\usepackage{comment}
\usepackage{enumitem}
\usepackage{flushend}
\usepackage{xspace}
\usepackage{pgfplots,tikz}
\pgfplotsset{compat=1.18}
\usepackage{amsthm}
\usepackage{thmtools,thm-restate}
\usepackage{hhline}
\usepackage{array}
\usepackage{epstopdf}  
\newtheorem{myDef}{Definition}

\def\BibTeX{{\rm B\kern-.05em{\sc i\kern-.025em b}\kern-.08em
    T\kern-.1667em\lower.7ex\hbox{E}\kern-.125emX}}
    
\begin{document}

\def\BY#1{\textcolor{red}{[Bin: #1]}}

\title{Data Driven Decision Making with Time Series and Spatio-temporal Data}

\author{\IEEEauthorblockN{Bin Yang}
\IEEEauthorblockA{\textit{East China Normal University} \\
Shanghai, China \\
byang@dase.ecnu.edu.cn}
\and
\IEEEauthorblockN{Yuxuan Liang}
\IEEEauthorblockA{\textit{Hong Kong University of} \\ \textit{Science and Technology (Guangzhou)} \\
Guangzhou, China \\
yuxliang@outlook.com}
\and
\IEEEauthorblockN{Chenjuan Guo}
\IEEEauthorblockA{\textit{East China Normal University} \\
Shanghai, China \\
cjguo@dase.ecnu.edu.cn}
\and
\IEEEauthorblockN{Christian S. Jensen}
\IEEEauthorblockA{\textit{Aalborg University} \\
Aalborg, Denmark \\
csj@cs.aau.dk}
}

\maketitle

\begin{abstract}
Time series data captures properties that change over time. Such data occurs widely, ranging from the scientific and medical domains to the industrial and environmental domains. When the properties in time series exhibit spatial variations, we often call the data spatio-temporal. As part of the continued digitalization of processes throughout society, increasingly large volumes of time series and spatio-temporal data are available. In this tutorial, we focus on data-driven decision making with such data, e.g., enabling greener and more efficient transportation based on traffic time series forecasting.

The tutorial adopts the holistic paradigm of ``data-governance-analytics-decision.'' We first introduce the data foundation of time series and spatio-temporal data, which is often heterogeneous. Next, we discuss data governance methods that aim to improve data quality. We then cover data analytics, focusing on the ``AGREE'' principles: Automation, Generalization,  Robustness, Explainability, and Efficiency. We finally cover data-driven decision making strategies and briefly discuss promising research directions. We hope that the tutorial will serve as a primary resource for researchers and practitioners who are interested in value creation from time series and spatio-temporal data. 
\end{abstract}


\section{Introduction}
As part of the continued digitalization of processes across society, increasingly large volumes of time series and spatio-temporal data are available. 
Managing and analyzing heterogeneous, multi-modal time-series and spatio-temporal data enables value creation by supporting data-driven decision-making across domains, such as smart cities, intelligent transportation, smart emergency response, digital energy, AIOps, smart agriculture,  ocean, and mining~\cite{DBLP:journals/sigmod/GuoJ014,li2024ocean,DBLP:journals/pvldb/PedersenYJ20}. 

As an example, path planning is a typical decision-making task in smart cities and intelligent transportation. Consider a scenario in which an autonomous taxi needs to take passengers to an airport. The taxi must select the most ``optimal'' route from a set of alternatives~\cite{DBLP:journals/vldb/YangDGJH18,guo2024efficient}. Data-driven decision-making facilitates a rational selection process: first, relevant \emph{multi-modal} traffic time series data, e.g., from road-side sensors and in-vehicle Global Navigation Satellite System capabilities, is collected. The quality of the collected raw data may be insufficient, thus requiring \emph{data governance} for enhancement. 
Subsequently, \emph{data analytics}, such as forecasting, are conducted on the data to predict future traffic conditions. Finally, the optimal route is chosen based on the predicted future traffic using \emph{data-driven decision making}, e.g., favoring the route with the highest probability of an on-time arrival. 
As another example,  consider cloud computing where resource scaling decisions must be made frequently. By gathering and managing historical cloud resource demand data, future demands can be predicted, particularly in the event of unexpected surges, allowing for timely resource auto-scaling to maintain service quality while minimizing energy consumption and carbon emissions~\cite{DBLP:journals/pvldb/PanWZY0CGWTDZYZ23}. 

\begin{figure}[!t]
	\begin{center}
		\includegraphics[scale=0.49]{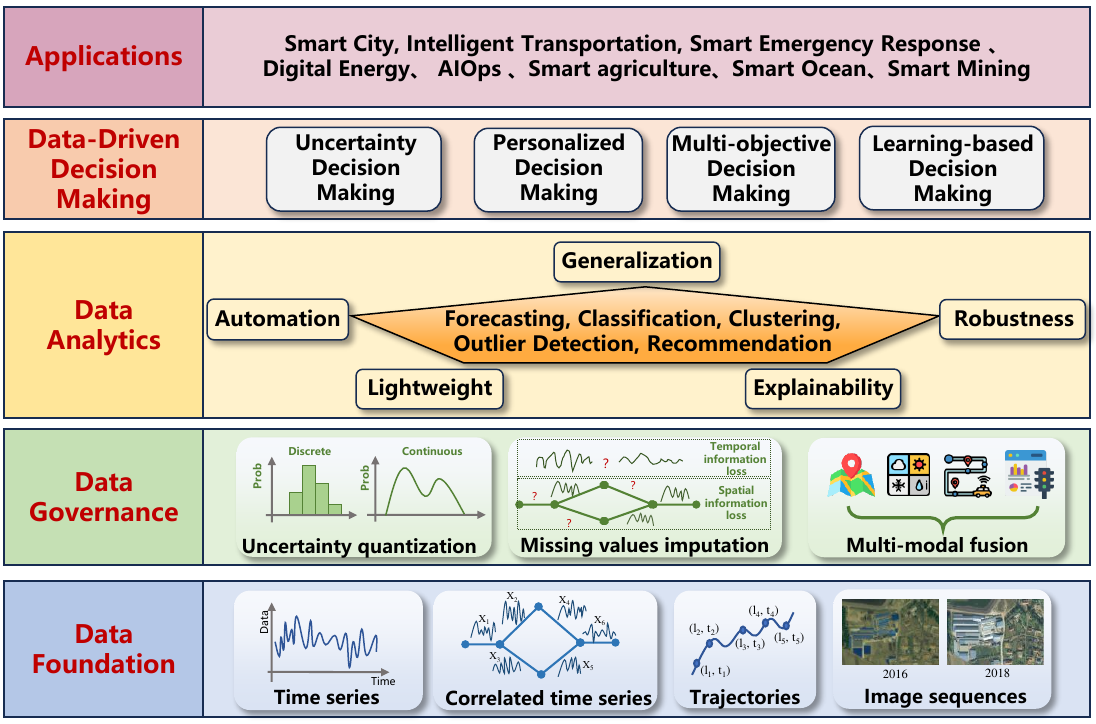}
	\end{center}
 \caption{Paradigm underlying Data Driven Decision Making---``Data-Governance-Analytics-Decision''}
	\label{fig:framework}
\end{figure}

We integrate the aspects of value creation from time series and spatio-temporal data into a data driven decision making paradigm with ‘\textbf{Data-Governance-Analytics-Decision}’, as depicted in Figure~\ref{fig:framework}.

\textit{Tutorial overview}. The tutorial is 1.5 hours long and is presented in a traditional lecture style. We summarize data foundations for multi-modal time series and spatio-temporal data, introduce data governance methods aiming at improving data quality, and cover data analytics methods, focusing on five characteristics: automation, generalization, robustness, explainability, and efficiency, which we summarize as the ``AGREE'' principles. We then cover various data-driven decision-making strategies, 
and finally, we briefly discuss research directions. 
The tutorial is organized as follows: (1) \textbf{Introduction and Data Foundation (5 mins)}. (2) \textbf{Data Governance (20 mins)}. (3) \textbf{Data Analytics (40 mins)}. (4) \textbf{Data-driven Decision Making (15 mins)}. (5) \textbf{Conclusion and Future Work (10 mins)}. 

\textit{Prior Offerings}: This tutorial is newly developed, which has not been presented in other venues. 

\textit{Target Audience}: The tutorial targets researchers specializing in database systems, data mining, machine learning, and data science, particularly those focused on time series and spatio-temporal data analysis. It is also relevant for practitioners engaged in real-world applications such as forecasting, classification, and anomaly detection related to smart cities, AIOps, Internet of Things, etc. The tutorial is data-centric, covering data governance, data analtyics, and data-driven decision making, thus well aligning with ICDE's core topics. The prerequisites include basic knowledge on databases, data mining, and deep learning. Expected benefits include: (1) Gain a deeper understanding of the ``data-governance-analytics-decision'' paradigm, covering advanced topics and recent progress on data governance, data analytics, and data-driven decision making with time seires and spaio-temporal data. (2) Interact with experts in this area and build connections that may lead to collaborations, mentorship, or career opportunities. (3) Understand emerging trends and technologies on time series and spatio-temporal data analytics. 

\textit{Related Tutorials}: 
We summarize related tutorials at recent top-tier venues. 
A tutorial at KDD'22~\cite{wen2022robust} 
focused on robust time series analysis. 
A tutorial at VLDB'23~\cite{keogh2023time} provided a concise selection of methods that can be integrated to effectively address a wide range of time series challenges. A tutorial at KDD'23~\cite{gong2023causal} focused on utilizing causal theory for 
temporal data mining. A tutorial at AAAI'24 
explore out-of-distribution generalization in time series. A tutorial at KDD'21~\cite{yu2021physics} provided an overview of 
physics-guided spatio-temporal data mining. The above tutorials focus on specific technical aspects in machine learning and data mining on time series and spatio-temporal data. In contrast, 
the present tutorial is data-centric and is based on the new holistic paradigm of ``data-governance-analytics-decision,'' encompassing data foundations, data governance methods, data analytics, and strategies for data-driven decision making.

\section{Tutorial Outline}

\subsection{Data Foundations}

Time series data captures dynamic properties over time, such as temperatures throughout the year. Spatio-temporal data combines time and location to characterize evolving events across temporal and spatial dimensions. Such data is often collected by different sensors, thus being multi-modal.

\begin{myDef}[\textbf{Time Series}] A time series captures properties over time and is given as $\mathbf{X}=\langle \mathbf{s}_1$, $\mathbf{s}_2$, $\dots$, $\mathbf{s}_M\rangle$, where $\mathbf{s}_i \in \mathbb{R}^C$, $1 \leq i \leq M$, is a $C$-dimensional vector, representing $C$ properties observed at the $i$-th timestamp.
\end{myDef}
\begin{myDef}[\textbf{Correlated Time Series}] A correlated time series is a set of $N$ interconnected time series, given as $\mathbf{T}=\{\mathbf{X}_1$, $\mathbf{X}_2$, $\dots$, $\mathbf{X}_N\}$. These time series often exhibit correlations due to the spatial arrangement of the corresponding sensors, and the correlations are often modeled using graphs. 

\end{myDef}
\begin{myDef}[\textbf{Trajectories}] Trajectories are sequences of (location, time) pairs that capture the positions of moving objects (e.g., humans, animals, and vehicles) over time.

\end{myDef}
\begin{myDef}[\textbf{Image Sequences}] An image sequence is given as $\mathbf{V}=\langle \mathbf{I}_1, \mathbf{I}_2,\dots ,\mathbf{I}_T \rangle$, where each $\mathbf{I}_i \in \mathbb{R}^{N\times M \times C}, 1 \leq i \leq T$, is an image that captures a state at the $i$-th timestamp. Each image comprises $N\times M$ pixels, representing spatial regions. Each pixel includes $C$ observed properties. 
\end{myDef}

\subsection{Data Governance}

As raw sensor data suffers from data quality issues, we cover effective data governance approaches that aim to enhance data quality.

\textbf{Missing Value Imputation}: Sensor data collection is prone to missing values due to sensor malfunctions or network disruptions. Within time series and spatio-temporal data management, missing value imputation is categorized into \textit{spatial completion, temporal completion}, and \textit{spatio-temporal completion} of missing data. 
Spatially, weighted graphs are often used to model sensor relationships, and spatially missing values are transformed into graph edge weight completion tasks. We consider graph-based semi-supervised learning~\cite{DBLP:journals/tkde/YangKJ14} and graph neural network autoencoders~\cite{DBLP:conf/icde/HuG0J19} for this purpose. Temporally, time series imputation and backcast techniques, such as recurrent neural network applications~\cite{Costa}, address missing temporal values. Spatio-temporally, integrating graph neural networks with recurrent neural networks~\cite{DBLP:conf/icde/Hu0GJX20} enables completion of missing values in Origin-Destination (OD) Matrix over time. In addition, spatio-temporally aware LLMs are employed to complete global significant wave heights using sparse buoy data~\cite{li2024ocean}.

\textbf{Uncertainty Quantification}: This process aims to capture uncertainty related to time and space using probability distributions.
Histograms or Gaussian mixture models are often used to represent distributions due to their capacity to approximate distributions without assumptions on the type of distribution. 
We cover a \textit{dynamic, uncertainty graph model} in the context of intelligent transportation. Traffic flow is inherently uncertain and dynamic, influenced by factors such as accidents and traffic volume. This can be modeled using $(I,D)$ pairs, where travel speed follows distribution $D$ over time interval $I$. We discuss two paradigms: the \textit{edge-centric paradigm}~\cite{DBLP:conf/icde/YangGJKS14} and \textit{path-centric paradigm}~\cite{DBLP:journals/vldb/YangDGJH18}. The former assigns distributions to edges, treating the edge distributions as independent, while the latter captures the distribution correlations along paths. Uncertainty quantification enhances data reliability, providing a robust foundation for further analytics and decision-making, e.g., decision making under uncertainty~\cite{DBLP:journals/pvldb/PanWZY0CGWTDZYZ23}.

\textbf{Multi-modal Fusion}: Time series and spatio-temporal data exhibit multi-source, heterogeneous, multi-modal characteristics, with significant correlations across time and space. Data fusion across modalities improves data quality by consolidating data, laying a robust foundation for further analysis and decision-making. Recent multi-modal fusion approaches can be summarized into four streams: \textit{feature-based, alignment-based, contrast-based, and generation-based} methods~\cite{zou2024deep}. A prime instance of alignment-based fusion is Map Matching~\cite{newson2009hidden}, which aligns vehicle trajectory data with corresponding map data. This process can eliminate trajectory noise as well as compensate for gaps in trajectories, thereby enhancing data quality and analytical and decision-making processes. Another example is feature-based fusion of multi-modal data, such as historical traffic, weather data, and point of interest,  for traffic forecasting~\cite{zhang2017deep,lin2019deepstn+}. Advances in representation learning and foundational models enable the capture of multi-modal spatio-temporal data as vectors or embeddings~\cite{jin2023time,yan2023urban,hao2024urbanvlp}. For example, MM-Path enhances path representations by aligning embeddings of the road network and satellite image of paths~\cite{xu2025mm}. 
By constructing temporal and spatial indices, we can improve the efficiency of analytics on diverse multi-modal spatio-temporal datasets.

\subsection{Data Analytics}

To facilitate data-driven decisions, enabling diverse analytics, e.g., forecasting, anomaly detection, and classification, on time-series and spatio-temporal data are essential. 
We do not cover specific analytics techniques; instead, we outline five desirable characteristics, which we call the \textbf{AGREE} principles--Automation, Generalization, Robustness, Explainability, and Efficiency. 

\textbf{Automation}: Traditional time series and spatio-temporal analysis methods are often crafted by human experts, which is resource-intensive. Automated methods offer a practical alternative: a search space is created for a specific task, upon which algorithms can identify and automatically assemble the most fitting models for the specific task and dataset. For instance, automated correlated time series forecasting automates the design of network architectures and hyperparameter selection~\cite{wupvldb,DBLP:journals/pacmmod/Wu0ZG0J23,pan2021autostg,DBLP:journals/vldb/WuWYZGQHSJ24,xinle2024FACTS}. 
Such automation not only enhances accuracy across diverse datasets and forecasting settings but also facilitates the discovery of optimal models that adhere to additional constraints, e.g., model sizes.

\textbf{Generalization}: Methods using deep learning are capable of superior performance in time-series and spatio-temporal analytics, 
but they depend on high-quality representations. Representation learning, which requires extensive labeled data, poses challenges in data acquisition and does not generalize across tasks~\cite{DBLP:journals/tkde/YangGY22}.  We consider general representation learning frameworks, pre-trained on abundant unlabeled data, that can be fine-tuned with limited labeled data for different spatio-temporal tasks. These include unsupervised contrastive learning~\cite{DBLP:conf/ijcai/YangGHT021}, weakly-supervised contrastive learning~\cite{DBLP:conf/icde/YangGHYTJ22},  masked autoencoders with relational inference~\cite{DBLP:conf/kdd/YangHGYJ23}, and cross-modal contrastive learning~\cite{xu2025mm}. 
For time series tasks, we propose general models for forecasting~\cite{wang2024rose} and anomaly detection~\cite{shentu2024towards} to eliminate the need for target-domain-specific training data.
These frameworks demonstrate generality and reduce the dependence on large amounts of labeled data. 
With the emergence of Artificial General Intelligence, we also consider how to leverage Large Language Models (LLMs) to enhance the zero-shot adaptability for time series forecasting~\cite{jin2023time,liu2023unitime} and spatio-temporal data mining~\cite{yan2023urban,hao2024urbanvlp}.

\textbf{Robustness}: Robustness in time-series and spatio-temporal analysis encompasses handling \textit{anomalous data}, \textit{data imbalance}, \textit{distribution shifts}, and \textit{multi-scale adaptation}. Traditional unsupervised anomaly detection algorithms assume implicitly that training occurs on fully-clean data, which is rarely available in practice. We thus cover robust anomaly detection~\cite{DBLP:conf/icde/KieuYGCZSJ22,DBLP:conf/icde/KieuYGJZHZ22, wu2025catch, contraAD} that operates effectively on noisy training data. Considering data imbalance, we cover an adversarial domain adaptation method~\cite{yunyaovldb24} that models unbalanced data patterns despite data size discrepancies. Addressing distribution shifts, e.g., due to changes in the physical world such as new roads, we cover continuous learning~\cite{haoicde24,kieu2024Team} tailored to spatio-temporal data dynamics. In addition, we cover means to model heterogeneous temporal patterns caused by temporal distribution shifts, e.g., in streaming data or non-stationary time series~\cite{QiuDEUT}. Lastly, we cover multi-scale modeling using pathways~\cite{pengiclr24} and ensemble learning strategies~\cite{DBLP:conf/ijcai/KieuYGJ19, davidpvldb} that enhance robustness by adaptively selecting and combining multiple scales for analytics.

\textbf{Explainability}: Effective decision-making requires both accuracy and explainability in data analytics. Although deep learning exhibits state-of-the-art accuracy at time series and spatio-temporal analyses, its ``black box'' nature hampers explainability. Considering this, we focus on two important aspects: {how to quantify the explainability of different methods} and {how to improve the explainability of deep learning methods}. We present a posthoc explainability metric~\cite{DBLP:conf/icde/KieuYGJZHZ22} for the assessment of autoencoder-based anomaly detection models. Furthermore, we consider how to leverage neural networks for feature extraction and integrate extracted features with interpretable models to augment overall prediction clarity~\cite{cheng2022pattern}. Additionally, we consider how to improve the explainability by tracking temporal associations among time series~\cite{razvanicde2021, DBLP:conf/icde/CirsteaYGKP22} and employing causal models to predict future correlations~\cite{kaivldb24}. We integrate physical equations into data-driven models to improve the explainability of air quality prediction~\cite{tian2024air-dualode}. Lastly, we cover how LLMs offer remarkable explainability in the realm of spatio-temporal data mining~\cite{yan2023urban,jin2023time}, enabling the injection of domain knowledge into language forms and facilitating the understanding of patterns within time series and spatio-temporal data.

\textbf{Efficiency}: 
Due to increasing privacy concerns, analytics are increasingly performed on personal edge devices rather than having to upload personal data to potentially untrustworthy servers. Contemporary analysis methods prioritize accuracy, resulting in the reliance on computationally intensive model architectures that are unsuitable for resource-constrained edge devices. Such methods also do not align with green computing principles. This calls for data analytics with high resource efficiency. To cover this aspect, we present LightPath~\cite{DBLP:conf/kdd/YangHGYJ23}, a knowledge distillation-based framework for efficient spatio-temporal data representation, and LightTS~\cite{DBLP:journals/pacmmod/0002Z0KGJ23}, an automated, lightweight temporal analysis framework capable of adapting quantization levels to align with memory limitations of edge devices. QCore~\cite{David2024Qcore} enables continuous calibrations for quantized models, addressing both resource efficiency and robustness on distribution shifts. EnhancingDFQ~\cite{enhancingDFQ} improves the accuracy of data-free model quantization by increasing sample-data diversity.
TimeDC studies a time series dataset condensation method that is able to compress large time series into a smaller counterpart while maintaining key properties~\cite{miao2024condensation}. This facilitates data analytics in resource constrained environments. 

Finally, as a wide variety of data analytics approaches have been proposed, it is essential to be able to compare such approaches empirically in a comprehensive and fair manner, thus calling for benchmarking. We will cover recent benchmarking on time series and spatio-temporal data analytics, focusing on forecasting and anomaly detection tasks implemented through end-to-end models~\cite{DBLP:journals/pvldb/PanWZY0CGWTDZYZ23} and foundation models~\cite{li2024foundts}. 

\subsection{Data-Driven Decision Making}
Decision making involves selecting among options. In our context, we consider temporal and spatial aspects while making decisions. 
Different scenarios require distinct strategies.  For scheduling, it is about optimizing resource use to meet demands efficiently and sustainably.  
Eco-driving focuses on reducing emissions through informed driving practices.  Predictive maintenance aims to preempt equipment failure to ensure uninterrupted operation. 
Instead of delving into the specifics of particular decision making scenarios, we provide an overview of data-driven strategies that are applicable to different decision making tasks.

\textbf{Decision Making under Uncertainty}: Decision making based on time-series and spatio-temporal data is often subject to uncertainty. This uncertainty can be quantified through data governance. Moreover, spatio-temporal analysis methods, such as predictive models, inherently capture uncertainty, typically using confidence intervals and probability distributions. At the same time, different stakeholders may have different risk preferences, including being risk-loving, risk-averse, and risk-neutral. 
By employing utility functions, we can encode different risk preferences, and then use expected utility to identify the most favorable options. We cover a novel pruning approach grounded in stochastic dominance~\cite{DBLP:journals/vldb/HuYGJ18, DBLP:journals/vldb/PedersenYJ20,DBLP:journals/tsas/PedersenYJM23}, enabling rapid identification of optimal choices across utility functions that encode different risk profiles.

\textbf{Multi-objective Decision Making}: Decision making often involves multiple criteria. For instance, cloud resource scheduling balances operational costs with quality of service (QoS), while fleet path planning considers travel distance and time, fuel consumption, and emissions. 
Multi-objective decision-making can be categorized into two classes: the first employs Pareto optimality to identify a set of non-dominated options, where no single option outperforms any other option across all objectives~\cite{DBLP:conf/icde/YangGJKS14}; the second consolidates multiple objectives into a single unified objective via a preference function, facilitating the selection of a single best option~\cite{DBLP:journals/vldb/0002GMJ15}.

\textbf{Personalized Decision Making}: This approach tailors decisions to individual preferences, which may include personalized risk profiles or preferences on multi-objective trade-offs. The challenge lies in selecting the most suitable preference for a given context—encompassing temporal, spatial, and other contexts like weather or events~\cite{DBLP:journals/tkde/YangGY22, DBLP:journals/vldb/GuoYHJC20}. Efficiently accommodating large numbers of preferences is critical. 

\textbf{Learning-based Decision Making}: These strategies simulate the decision patterns of experts, enabling non-experts to replicate expert-level decisions. For instance, we mimic expert taxi drivers' routing strategies by extracting valuable knowledge from their historical trajectories and decision patterns~\cite{DBLP:conf/icde/Guo0HJ18}. HiSSD~\cite{liu2025learning} incorporates cooperation temporal learning within a hierarchical architecture, identifying shared and specific skills across tasks, thereby enhancing policy generalization.


\subsection{Research Directions}

Most existing studies are highly specialized, focusing on individual cases, specific locations, and time. 
This specificity leads to models with limited applicability, high costs, and reduced scalability. To broaden the technology’s reach and reduce barriers to its use, the future of temporal and spatio-temporal decision intelligence may include: (1) Pre-training and few-shot learning: Leveraging pre-training on diverse tasks and extensive datasets across various spatio-temporal contexts can yield foundational models with broad applicability. Such models may facilitate few-shot or zero-shot learning when applied to novel tasks and domains.
(2) Generative model potential: Advances in generative models may improve performance in temporal and spatio-temporal decision-making. The strong generalization capabilities and precision in data generation of such models could enable better spatio-temporal transferability.
(3) Integration with LLMs: Merging temporal and spatio-temporal decision-making with LLMs may enable more intuitive and seamless interactions through natural language.

\section{Presenters} 

\noindent
\textbf{Bin Yang} is a Chair Professor at East China Normal University, working on data-driven decision intelligence, with a focus on time series and spatio-temporal data. He has published extensively on the tutorial's topics in database and data mining venues, e.g., ICDE, PVLDB, SIGMOD, and KDD, and machine learning venues, e.g., NeruIPS, ICLR, and IJCAI.

\noindent
\textbf{Yuxuan Liang} is a tenure-track Assistant Professor at Hong Kong University of Science and Technology (Guangzhou), working on spatio-temporal data mining and urban computing. He has published 70+ papers in prestigious venues (e.g., TPAMI, AI, TKDE, KDD, NeurIPS, and WWW). 


\noindent
\textbf{Chenjuan Guo} is a Professor at East China Normal University, working on AI for time series and spatio-temporal data, in particular, on pretrained foundation models for time series and spatio-temporal data.  

\noindent
\textbf{Christian S. Jensen} is a Professor at Aalborg University, a member of the Academia Europaea, and an ACM and IEEE fellow, working on spatio-temporal data management and analytics. He received ACM SIGMOD Contributions Award and IEEE TCDE Impact Award.

\bibliographystyle{IEEEtran}
\bibliography{main}

\end{document}